\documentstyle[multicol,aps,epsfig,amssymb]{revtex}

\begin{document}

\title{Optimizing Epochal Evolutionary Search:\\
Population-Size Dependent Theory}
\author{Erik van Nimwegen$^{\dag\ddag}$ and James P. Crutchfield$^\dag$\\
$\dag$ Santa Fe Institute, 1399 Hyde Park Road, Santa Fe, NM 87501\\
$\ddag$ Theoretical Biology and Bioinformatics Group, University of Utrecht,\\
Padualaan 8, NL-3584-CH Utrecht, The Netherlands\\ 
\{erik,chaos\}@santafe.edu}

\maketitle


\begin{abstract}
Epochal dynamics, in which long periods of stasis in an evolving
population are punctuated by a sudden burst of change, is a common
behavior in both natural and artificial evolutionary processes. We
analyze the population dynamics for a class of fitness functions that
exhibit epochal behavior using a mathematical framework developed
recently. In the latter the approximations employed led to a
population-size independent theory that allowed us to determine
optimal mutation rates. Here we extend this approach to include the
destabilization of epochs due to finite-population fluctuations and
show that this dynamical behavior often occurs around the optimal
parameter settings for efficient search. The resulting, more accurate
theory predicts the total number of fitness function evaluations to
reach the global optimum as a function of mutation rate, population
size, and the parameters specifying the fitness function. We further
identify a generalized error threshold, smoothly bounding the
two-dimensional regime of mutation rates and population sizes for
which evolutionary search operates efficiently.

\begin{center}
Santa Fe Institute Working Paper 98-10-090
\end{center}
\end{abstract}


\begin{multicols}{2}

\newlength{\Temp}
\setlength{\Temp}{\parskip}
\setlength{\parskip}{-0.03in}
\tableofcontents
\setlength{\parskip}{\Temp}


\section{Designing Evolutionary Search}

Evolutionary search algorithms are a class of stochastic optimization
procedures inspired by biological evolution, e.g see Refs.
\cite{Back96a,Koza93a,Goldberg89c,Mitchell96a}: a population of
candidate solutions evolves under selection and random ``genetic''
diversification operators. Evolutionary search algorithms have been
successfully applied to a diverse variety of optimization problems;
see, for example Refs. \cite{ICGA91,Cham95a,Davis91a,ICGA95,ICGA93}
and references therein. Unfortunately, and in spite of a fair amount
of theoretical investigation, the mechanisms constraining and driving
the dynamics of evolutionary search on a given problem are often not
well understood.

There are very natural difficulties that are responsible for this
situation. In mathematical terms evolutionary search algorithms are
population-based discrete stochastic nonlinear dynamical systems. In
general, the constituents of the search problem, such as the structure
of the fitness function, selection, finite-population fluctuations,
and genetic operators, interact in complicated ways to produce a rich
variety of dynamical behaviors that cannot be easily understood in
terms of the constituents individually. These complications make a
strictly empirical approach to the question of whether and how to use
evolutionary search problematic.

The wide range of behaviors exhibited by nonlinear population-based
dynamical systems have been appreciated for decades in the field of
mathematical population genetics. Unfortunately, this appreciation has
not led to a quantitative predictive theory that is applicable to the
problems of evolutionary search; something desired, if not required,
for the engineering use of stochastic search methods.

We believe that a general, predictive theory of the dynamics of
evolutionary search can be built incrementally, starting with a
quantitative analytical understanding of specific problems and then
generalizing to more complex situations. In this vein, the work
presented here continues an attempt to unify and extend theoretical
work in the areas of evolutionary search theory, molecular evolution
theory, and mathematical population genetics. Our strategy is to
focus on a simple class of problems that, nonetheless, exhibit some
of the rich behaviors encountered in the dynamics of evolutionary search
algorithms. Using analytical tools from statistical mechanics, dynamical
systems theory, and the
above mentioned fields we developed a detailed and quantitative
understanding of the search dynamics for a class of problems that
exhibit epochal evolution. On the one hand, this allows us to
analytically predict optimal parameter settings for this class of
problems. On the other hand, the detailed understanding of the
behavior for this class of problems provides valuable insights into
the emergent mechanisms that control the dynamics in more general
settings of evolutionary search and in other population-based dynamical
systems.

In a previous paper, Ref. \cite{Nimw98a}, we showed how a detailed
dynamical understanding, announced in Ref. \cite{Nimw97a} and expanded
in Ref. \cite{Nimw97b}, can be turned to practical advantage.
Specifically, we determined how to set the mutation rate to reach, in
the fewest number of fitness function evaluations, the global optimum
in a wide class of fitness functions. Due to certain cancellations at
the level of approximation used, the resulting theory lead to
population-size independent predictions.

Here we extend this theory to include additional important effects, such
as the increased search effort caused by the dynamical destabilization
of epochs, to be explained below, which reintroduce the dependence on
population size. The result is a more accurate theory that analytically
predicts the total number of fitness function evaluations needed on
average for the algorithm to discover the global optimum of the fitness
function.

In addition, we develop a detailed understanding of the operating regime in
parameter space for which the search is performed most efficiently. We believe
this will provide useful guidance on how to set search algorithm parameters
for more complex problems. In particular, our theory explains the marginally
stable behavior of the dynamics when the parameters are set to minimize
search effort. Most simply put, the optimal parameter setting occurs when the
dynamics is as stochastic as possible without corrupting information
stored in the population about the location of the current best genotypes.
The results raise the general question of whether it is desirable for optimal
search to run in dynamical regimes that are a balance of stability and
instability. The mechanisms we identify suggest how this balance is, in
fact, useful.


\section{Royal Staircase Fitness Functions}

Choosing a class of fitness functions to analyze is a delicate
compromise between generality, mathematical tractability, and the
degree to which the class is representative of problems often
encountered in evolutionary search. A detailed knowledge of the
fitness function is very {\em atypical} of evolutionary search
problems. If one knew the fitness function in detail, one would not
have to run an evolutionary search algorithm to find high fitness
solutions in the first place. The other extreme of assuming complete
generality, however, cannot lead to enlightening results either, since
averaged over all problems, all optimization algorithms perform
equally well (or badly); see Ref. \cite{Wolp97a}. We thus focus on a
specific subset of fitness functions, somewhere between these
extremes, that we believe at least have ingredients typically
encountered in evolutionary search problems and that exhibit widely
observed dynamical behaviors in both natural and artificial
evolutionary processes.

In our preceding paper, Ref. \cite{Nimw98a}, we justified in some
detail our particular choice of fitness function both in terms of
biological motivations and in terms of optimization engineering
issues. In short, many biological systems and optimization problems
have highly degenerate genotype-to-phenotype maps; that is, the
mapping from genetic specification to fitness is a many-to-one
function. Consequently, the number of different fitness values that
genotypes can take is much smaller than the number of different
genotypes.

Additionally, due to the many-to-one mapping and since genotype spaces
are generally of very high dimensionality, the genotype space tends to
break into networks of ``connected'' sets of equal-fitness genotypes
that can reach each other via elementary genetic variation steps such
as point mutations. These connected subsets of isofitness genotypes
are generally referred to as ``neutral networks'' in molecular
evolution theory, see Refs.
\cite{Font98a,Huynen&Stadler&Fontana,Huynen95,Reidys98b,Weber97}. This
leads us to posit that the genotype space for general search problems
decomposes into a number of such neutral networks. We also assume that
higher fitness networks are smaller in volume than low fitness
networks. Finally, we assume that from any neutral network there
exist connections to higher fitness networks such that, taken as a
whole, the fitness landscape has no local optima other than the global
optimum.

Under these assumptions, genotype space takes on a particular type of
architecture: ``subbasins'' of the neutral networks are connected by
``portals'' leading between them and so to higher or lower fitness.
Stated in the simplest terms possible, the evolutionary population
dynamics then becomes a type of diffusion constrained by this
architecture. For example, individuals in a population diffuse over
neutral networks until a portal to a network of higher fitness is
discovered and the population moves onto this network.

In order to model the behavior associated with the subbasin-portal
architecture, we defined the class of {\em Royal Staircase} fitness
functions that capture the essential elements sketched above.
Importantly, this class of fitness functions is simple enough to
admit a fairly detailed quantitative mathematical analysis of the
associated epochal evolutionary dynamics.

The Royal Staircase fitness functions are defined as follows.
\begin{enumerate}
\item{Genomes are specified by binary strings
        $s = s_1 s_2 \cdots s_L, s_i \in \{0,1\},$ of length $L = N K$.}
\item{Reading the genome from left to right, the number $I(s)$ of
        consecutive $1$s is counted.}
\item{The fitness $f(s)$ of string $s$ with $I(s)$ consecutive ones,
        followed by a zero, is $f(s) = 1 + \lfloor I(s) / K \rfloor$. The
        fitness is thus an integer between $1$ and $N+1$.}
\end{enumerate}
Four observations are in order.
\begin{enumerate}
\item{The fitness function has two parameters, the number $N$ of blocks
	and the number $K$ of bits per block. Fixing them determines
    a particular optimization problem or fitness ``landscape''.}
\item{There is a single global optimum: the genome $s = 1^L$---namely,
    the string of all $1$s---with fitness $f(s) = N+1$.}
\item{The proportion $\rho_n$ of genotype space filled by strings of
    fitness $n$ is given by:
\begin{equation}
\rho_n = 2^{-K(n-1)} \left( 1- 2^{-K}\right),
\label{nFitStrings}
\end{equation}
for $n \leq N$. Thus, high fitness strings are exponentially more rare
than low fitness strings.}
\item{For each block of $K$ bits, the all-$1$s pattern is the one that
    confers increased fitness on a string. Without loss of generality,
    any of the other $2^K-1$ configurations could have been chosen as
    the ``correct'' configuration, including different patterns for
    each of the $N$ blocks. Furthermore, since the GA here does not
    use crossover, arbitrary permutations of the $L$ bits in the
    fitness function definition leave the evolutionary dynamics
    unchanged.}
\end{enumerate}
The net result is that the Royal Staircase fitness functions implement
the intuitive idea that increasing fitness is obtained by setting more
and more bits in the genome ``correctly''. One can only set correct bit
values in sets of $K$ bits at a time, creating an ``aligned'' block, and
in blocks from left to right. A genome's fitness is proportional to the
number of such aligned blocks. And since the $(n+1)$st block only confers
fitness when all $n$ previous blocks are aligned as well, there is
contingency between blocks. This realizes our view of the underlying
architecture as a set of isofitness genomes that occur in nested neutral
networks of smaller and smaller volume. (Cf. Figs. 1 and 2 of Ref.
\cite{Nimw98a}.)

\section{The Genetic Algorithm}

For our analysis of evolutionary search we have chosen a simplified form
of a genetic algorithm (GA) that does not include crossover and that
uses fitness-proportionate selection. The GA is defined by
the following steps.
\begin{enumerate}
\item{Generate a population of $M$ bit strings of length $L = NK$ with
    uniform probability over the space of $L$-bit strings.}
\item{Evaluate the fitness of all strings in the population.}
\item{Stop, noting the generation number $t_{\rm opt}$, if a string with
        optimal fitness $N+1$ occurs in the population. Else, proceed.}
\item{Create a new population of $M$ strings by selecting, with
    replacement and in proportion to fitness, strings from the current
    population.}
\item{Mutate, i.e. change, each bit in each string of the new
    population with probability $q$.}
\item{Go to step 2.}
\end{enumerate}
When the algorithm terminates there have been $E = M t_{\rm opt}$
fitness function evaluations.

In Ref. \cite{Nimw98a} we motivated our excluding crossover and
discussed at some length the reasons that crossover's role in epochal
evolution is not expected to be significant due to population
convergence effects.

This GA effectively has two parameters: the mutation rate $q$ and the
population size $M$. A given optimization problem is specified by the
fitness function in terms of $N$ and $K$. Stated most prosaically, then,
the central goal of the following analysis is to find those settings
of $M$ and $q$ that minimize the average number $\langle E\rangle$ of
fitness function queries for given $N$ and $K$ required to discover the
global optimum. Our approach is to develop analytical expressions for
$E$ as a function of $N$, $K$, $M$, and $q$ and then to study the
{\em search-effort surface} $E(q,M)$ at fixed $N$ and $K$. Before beginning the analysis,
however, it is helpful to develop an appreciation of the basic dynamical
phenomenology of evolutionary search on this class of fitness functions.
Then we will be in a position to lay out the evolutionary equations of
motion and analyze them.

\section{Observed Population Dynamics}

The typical behavior of a population evolving on a fitness ``landscape''
of connected neutral networks, such as defined above, alternates between
long periods ({\em epochs}) of stasis in the population's average fitness
and sudden increases ({\em innovations}) in the average fitness. (See, for
example, Fig. 1 of Ref. \cite{Nimw97b} and Fig. 1 of Ref. \cite{Nimw98a}.)

We now briefly recount the experimentally observed behavior of typical
Royal Staircase GA runs in which the parameters $q$ and $M$ are set
close to their optimal setting. The reader is referred to Ref.
\cite{Nimw97b} for a detailed discussion of the dynamical regimes this
type of GA exhibits over a range of different parameter settings.

\end{multicols}

\begin{figure}[htbp]
  \centerline{\epsfig{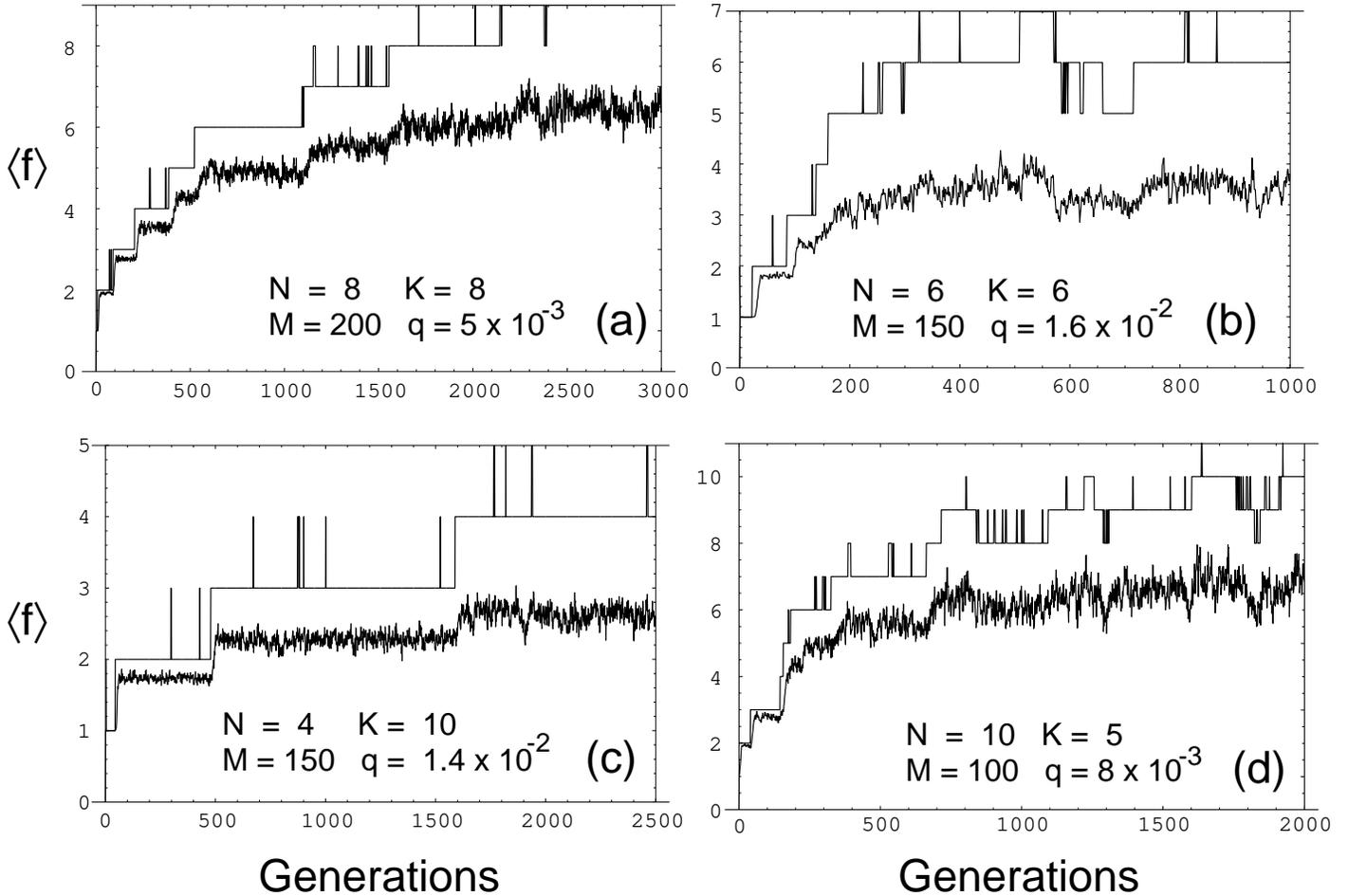}}
\caption{Examples of the Royal Staircase GA population dynamics with
  different parameter settings. The four plots show best fitness in
  the population (upper lines) and average fitness in the population
  (lower lines) as a function of time, measured in generations. The
  fitness function and GA parameters are given in each plot. In each
  case we have chosen $q$ and $M$ in the neighborhood of their optimal
  settings (see later) for each of the four values of $N$ and $K$.} 
\label{fig_runs}
\end{figure}

\begin{multicols}{2}
  
Figure \ref{fig_runs} illustrates the GA's behavior at four different
parameter settings. Each individual figure plots the best fitness in the population
(upper lines) and the average fitness $\langle f \rangle$ in the population
(lower lines) as a function of the number of generations. Each plot
is produced from a single GA run. In all of these runs the average
fitness $\langle f \rangle$ in the population goes through stepwise
changes early in the run, alternating epochs of stasis with sudden
innovations in fitness. Later in each run, especially for those in
Figs. \ref{fig_runs}(b) and \ref{fig_runs}(d), $\langle f \rangle$
tends to have higher fluctuations and the epochal nature of the
dynamics becomes unclear.

In the GA runs the population starts out with strings that only have
relatively low fitness, say fitness $n$ (in all four plots of Fig.
\ref{fig_runs} we have $n=1$). Selection and mutation then
establish an equilibrium in the population until a string aligns the
$n$th block, and descendants of this string with fitness $n+1$ spread
through the population. A new equilibrium is then established until a
string of fitness $n+2$ is discovered and so on, until finally, a
string of fitness $N+1$ is discovered.

Notice that $\langle f \rangle$ roughly tracks the epochal behavior of
the best fitness in the population. Every time a newly discovered
higher fitness string has spread through the population,
$\langle f \rangle$ reaches a new, higher equilibrium value around which it
fluctuates. As a run progresses to higher epochs, $\langle f \rangle$
tends to have higher fluctuations and the epochal nature of the
dynamics is obscured. This is a result of the fact that for the
highest epochs the difference between $\langle f \rangle$ in consecutive
epochs is smaller than the average fitness fluctuations induced by
the finite-population sampling; see Ref. \cite{Nimw97b}.

Notice, too, that often the best fitness shows a series of brief
jumps to higher fitness during an epoch. When this occurs strings
of higher fitness are discovered but, rather than spreading through
the population, are lost within a few generations.

For each of the four settings of $N$ and $K$ we have
chosen the values of $q$ and $M$ such that the average total
number $\langle E\rangle $ of fitness function evaluations to
reach the global optimum for the first time is minimal. Thus, the
four plots illustrate the GA's typical dynamics close to optimal
$(q,M)$-parameter settings.

Despite what appears at first blush to be relatively small variations
in fitness function and GA parameters, there is a large range,
almost a factor of $10$, in times to reach the global optimum across
the runs. Thus, there can be a strong parameter dependence in search
times. It also turns out that the standard deviation $\sigma$ of the
mean total number $\langle E \rangle$ of fitness function evaluations
is of the same order as $\langle E \rangle$. (See Table
\ref{FitEvalsTable}.) Thus, there are large run-to-run variations in
the time to reach the global optimum. This is true for all parameter
settings with which we experimented, of which only a few are reported
here.

Having addressed the commonalities between runs, we now turn to
additional features that each illustrates.
Figure \ref{fig_runs}(a) shows the results of a GA run with $N=8$
blocks of $K=8$ bits each, a mutation rate of $q=0.005$, and a
population size of $M=200$. During the epochs, the best
fitness in the population hops up and down several times before it
finally jumps up and the new more-fit strings stabilize in the
population. This transition is reflected in the average fitness
also starting to move upward. In this particular run, it took the
GA approximately $3.4 \times 10^5$ fitness function evaluations (1700
generations) to discover the global optimum for the first time.
Over 500 runs, the GA takes on average $5.3 \times 10^5$ fitness function
evaluations to reach the global optimum for these parameters. The
inherent large per-run variation means in this case that some runs
take less than $10^5$ function evaluations and that others take many
more than $10^6$.

Figure \ref{fig_runs}(b) plots a run with $N=6$ blocks of length $K=6$
bits, a mutation rate of $q=0.016$, and a population size of $M=150$.
The GA discovered the global optimum after approximately $4.8 \times
10^4$ fitness function evaluations (325 generations). For these
parameters, the GA uses approximately $5.5 \times 10^4$ fitness
function evaluations on average to reach the global fitness optimum.
Notice that the global optimum is only consistently present in the
population between generations $530$ generation $570$. After that, the
global optimum is lost again until after generation $800$. As we will
show, this is a typical feature of the GA's behavior for parameter
settings close to those that give minimal $\langle E \rangle$. The
global fitness optimum often only occurs in relatively short bursts
after which it is lost again from the population. Notice also that
there is only a small difference in $\langle f \rangle$ depending
whether the best fitness is either $6$ or $7$ (the optimum).

Figure \ref{fig_runs}(c) shows a run for a small number ($N=4$) of
large ($K=10$) blocks. The mutation rate is $q=0.014$ and the
population size is again $M=150$. As in the three other runs we see
that $\langle f \rangle$ goes through epochs punctuated by rapid
increases in $\langle f \rangle$. We also see that the best fitness in the
population jumps several times before the population fixes on a higher
fitness string. The GA takes about $1.9 \times 10^5$ fitness function
evaluations on average to discover the global optimum for these
parameter settings. In this run, the GA first discovered the global
optimum after $2.7 \times 10^5$ fitness function evaluations. Notice
that the optimum never stabilized in the population.

Finally, Fig. \ref{fig_runs}(d) shows a run with a large number
($N=10$) of small ($K=5$) blocks. The mutation rate is $q=0.008$ and
the population size is $M=100$. Notice that in this run, the best
fitness in the population alternates several times between fitnesses
$8$, $9$, and $10$ before it reaches (fleetingly) the global fitness
optimum of $11$. Quickly after it has discovered the global optimum,
it disappears again and the best fitness in the population largely
alternates between $9$ and $10$ from then on. It is notable that this
intermittent behavior of the best fitness is barely discernible in the
behavior of $\langle f \rangle$. It appears to be lost in the
``noise'' of the average fitness fluctuations. The GA takes about $1.2
\times 10^5$ fitness function evaluations on average at these
parameter settings to reach the global optimum; while in this
particular run the GA took $1.6 \times 10^5$ fitness function
evaluations (1640 generations) to briefly reach the optimum for the
first time.

\section{Statistical Dynamics of Evolutionary Search}

In Refs. \cite{Nimw97a} and \cite{Nimw97b} we developed the
statistical dynamics of genetic algorithms to analyze the behavioral
regimes of a GA searching the Royal Road fitness functions, which are
closely related to the Royal Staircase fitness functions just defined.
The analysis here builds on those results and, additionally, is a
direct extension of the optimization analysis and calculations in Ref.
\cite{Nimw98a}. We briefly review the essential points from these
previous papers. We refer the reader to Ref. \cite{Nimw97b} for a
detailed description of the similarities and differences of our
theoretical approach with other theoretical approaches such as the
work by Pru\"{g}el-Bennett, Rattray, and Shapiro, Refs.
\cite{PrugelBennett&Shapiro94,PrugelBennett&Shapiro96,Rattray&Shapiro96},
the diffusion equation methods developed by Kimura,
Refs. \cite{Kimura64,Kimurabook}, and the quasispecies theory,
Ref. \cite{Eigen&McCasKill&Schuster89}.

\renewcommand{\arraystretch}{1.1}
\begin{table}[htbp]
  \begin{center}
  \begin{tabular}{||c|c|c|c|c|c||}
  $N$ & $K$ & $M$ & $q$ & $\langle E \rangle$ & $\sigma$ \\ \cline{1-6}
  8 & 8 & 200 & $0.005$ &  $5.3 \times 10^5$ & $2.1 \times 10^5$ \\ \cline{1-6} 
  6 & 6 & 150 & $0.016$ &  $5.5 \times 10^4$ & $3.0 \times 10^4$ \\ \cline{1-6} 
  4 & 10 & 150 & $0.014$ &  $1.9 \times 10^5$ & $1.0 \times 10^5$ \\ \cline{1-6} 
  10 & 5 & 100 & $0.008$ &  $1.2 \times 10^5$ & $4.9 \times 10^4$
  \end{tabular}
  \end{center}
\caption{Mean $\langle E \rangle$ and standard deviations $\sigma$
  of the expected number of fitness function evaluations for the Royal
  Staircase fitness functions and GA parameters shown in the runs of
  Fig. \ref{fig_runs}. The estimates were made from 500 GA runs.
  }
\label{FitEvalsTable}
\end{table}

\subsection{Macrostate Space}

Formally, the state of a population in an evolutionary search
algorithm is only specified when the frequency of occurrence of each
of the $2^L$ genotypes is given. Thus, the dimension of the
corresponding microscopic state space is very large. One immediate
consequence is that the evolutionary dynamic, on this level, is given
by a stochastic (Markovian) operator of size ${\cal O}(2^L \times 2^L)$.
Generally, using such a microscopic description makes
analytical and quantitative predictions of the GA's behavior
unwieldy. Moreover, since the practitioner is generally
interested in the dynamics of some more macroscopic statistics, such as
best and average fitness, a microscopic description is uninformative
unless an appropriate projection onto the desired macroscopic
statistic is found.

With these difficulties in mind, we choose to describe the macroscopic
state of the population by its fitness distribution, denoted by a
vector $\vec{P} = (P_1 , P_2 , \ldots , P_{N+1} )$, where the
components $0 \leq P_f \leq 1$ are the proportions of individuals in
the population with fitness $f = 1, 2, \ldots, N+1$. We refer to
$\vec{P}$ as the {\em phenotypic quasispecies}, following its analog
in molecular evolution theory; see Refs.
\cite{Eigen71,Eigen&McCasKill&Schuster89,Eigen&Schuster77}. Since
$\vec{P}$ is a distribution, it is normalized:
\begin{equation}
\sum_{f=1}^{N+1} P_f = 1.
\label{Normalization}
\end{equation}
The average fitness $\langle f \rangle$ of the population is given by:
\begin{equation}
\langle f \rangle = \sum_{f=1}^{N+1} f P_f.
\label{AvFitness}
\end{equation}

\subsection{The Evolutionary Dynamic}

The fitness distribution $\vec{P}$ does not uniquely specify the
microscopic state of the population; that is, there are many
microstates with the same fitness distribution. An essential
ingredient of the statistical dynamics approach is to assume a maximum
entropy distribution of microstates conditioned on the macroscopic
fitness distribution. Note that our approach shares a focus on fitness
distributions and maximum entropy methods with that of
Pru\"{g}el-Bennett, Rattray, and Shapiro, Refs.
\cite{PrugelBennett&Shapiro94,PrugelBennett&Shapiro96,Rattray&Shapiro96}.
In our case, the maximum entropy assumption entails that, given a
fitness distribution $\vec{P}(t)$ at generation $t$, each microscopic
population state with this fitness distribution is equally likely to
occur. Given this assumption, we can construct a generation operator
${\bf G}$ that acts on the current fitness distribution to give the
{\em expected} fitness distribution of the population at the next time
step. (See $\vec{P}(t) \rightarrow {\bf G} [ \vec{P}(t)]$ illustrated
in Fig. \ref{SamplingDynamics}.) In the limit of infinite
populations, which is similar to the thermodynamic limit in
statistical mechanics, this operator ${\bf G}$ maps the current
fitness distribution $\vec{P}(t)$ deterministically to the fitness
distribution $\vec{P}(t+1)$ at the next time step; that is,
\begin{equation}
\vec{P}(t+1) = {\bf G} [ \vec{P}(t) ] ~.
\end{equation}
Simulations indicate that for very large populations ($M \gtrapprox
2^L$) the dynamics on the level of fitness distributions is indeed
deterministic and given by the above equation; thereby justifying
the maximum entropy assumption in this limit.

The operator ${\bf G}$ consists of a selection operator ${\bf S}$
and a mutation operator ${\bf M}$:
\begin{equation}
{\bf G} = {\bf M} \cdot {\bf S}.
\end{equation}
The selection operator encodes the fitness-level effect of selection on
the population; and the mutation operator, the fitness-level effect of
mutation. Appendixes \ref{SelectionOperator} and \ref{MutationOperator}
review the construction of these operators for our GA and the Royal
Staircase fitness functions.

For now, we note that the infinite population dynamics can be obtained
by iteratively applying the operator ${\bf G}$ to the initial fitness
distribution $\vec{P}(0)$. Thus, the solutions to the macroscopic
equations of motion, in the limit of infinite populations, are
formally given by
\begin{equation}
\vec{P}(t) = {\bf G}^{(t)} [ \vec{P}(0) ] ~.
\label{InfPopEqoMo}
\end{equation}
Recalling Eq. (\ref{nFitStrings}), it is easy to see that the initial
fitness distribution $\vec{P}(0)$
is given by:
\begin{equation}
\label{init_fit_dist}
P_n(0) = 2^{-K (n-1)} \left( 1- 2^{-K}\right) \;, ~ 1 \leq n \leq N ~,
\end{equation}
and
\begin{equation}
P_{N+1}(0) = 2^{-K N}.
\label{init_fit_dist_0}
\end{equation}
As shown in Refs. \cite{Nimw97a} and \cite{Nimw97b}, despite ${\bf G}$'s
nonlinearity, it can be linearized such that the $t$th iterate
${\bf G}^{(t)}$ can be directly obtained by solving for the eigenvalues
and eigenvectors of the linearized version ${\bf \tilde{G}}$. This
leads to a closed-form solution of the infinite-population dynamics
specified by Eq. (\ref{InfPopEqoMo}).

\subsection{Finite Population Sampling}

For large ($M \gtrapprox 2^L$) and infinite populations the dynamics
of the fitness distribution is qualitatively very different from the
behavior shown in Fig. \ref{fig_runs}: $\langle f \rangle$ increases
smoothly and monotonically to an asymptote over a small number of
generations. That is, there are no epochs. The reason is that for an
infinite population, all genotypes are present in the initial
population. Instead of the evolutionary dynamics {\em discovering}
fitter strings over time, it essentially only expands the proportion
of globally optimal strings already present in the initial population
at $t=0$. In spite of the qualitatively different dynamics for large
populations, we showed in Ref. \cite{Nimw97b} that the (infinite
population) operator ${\bf G}$
is the essential ingredient for describing the finite-population
dynamics with its epochal dynamics as well.

There are two important differences between the infinite-population
dynamics and that with finite populations. The first is that with finite
populations the components $P_n$ cannot take on continuous values
between $0$ and $1$. Since the number of individuals with fitness $n$ in
the population is necessarily an integer, the values of $P_n$ are
quantized in multiples of $1/M$. Thus, the space of allowed finite
population fitness distributions turns into a regular lattice in $N+1$
dimensions with a lattice spacing of $1/M$ within the simplex specified
by normalization (Eq. (\ref{Normalization})).

Second, due to the sampling
of members in the finite population, the dynamics of the fitness
distribution is no longer deterministic. In general, we can only
determine the conditional probabilities ${\rm Pr}[\vec{Q}|\vec{P}]$
that a given fitness distribution $\vec{P}$ leads to another
$\vec{Q} = (Q_1, \ldots , Q_{N+1})$ in the next generation.

\begin{figure}[htbp]
\centerline{\epsfig{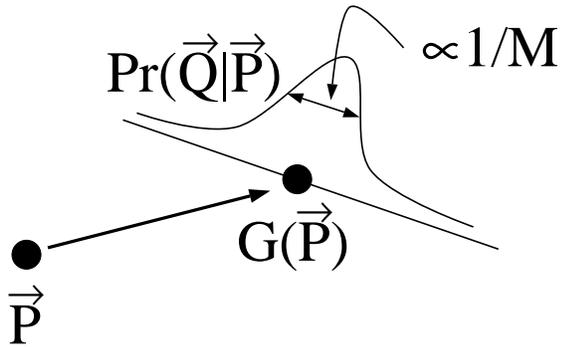}}
\caption{Illustration of the stochastic dynamics involved in going from
  one generation to the next starting with finite population $\vec{P}$,
  moving to the next (expected) population ${\bf G} [\vec{P}]$, and then
  sampling to obtain the distribution ${\rm Pr}[\vec{Q}|\vec{P}]$ of
  finite populations at the next time. The width of the latter
  distribution is inversely proportional to the population size $M$.
  Note that the underlying state space is a discrete lattice
  with spacing $1/M$.}
\label{SamplingDynamics}
\end{figure}

It turns out that the probabilities ${\rm Pr}[\vec{Q}|\vec{P}]$ are
given by a multinomial distribution with mean ${\bf G} [ \vec{P} ]$:
\begin{equation}
{\rm Pr} [ \vec{Q} | \vec{P} ] = M!
\prod_{n=1}^{N+1} \frac{\left({\bf G}_n [ \vec{P} ]
  \right)^{m_n}}{m_n!} ~.
\label{SamplingDist}
\end{equation}
where $Q_n = m_n/M$, with $0 \leq m_n \leq M$ integers. (The stochastic
effects of finite sampling are illustrated in Fig.
\ref{SamplingDynamics}.) For any finite-population fitness distribution
$\vec{P}$ the (infinite population) operator ${\bf G}$ gives the GA's
{\em average} dynamics over one time step, since by Eq.
(\ref{SamplingDist}) the {\em expected}
fitness distribution at the next time step is ${\bf G} [ \vec{P} ]$.
Note that the components ${\bf G}_n [ \vec{P} ]$ need not be multiples
of $1/M$. Therefore, the {\em actual} fitness distribution $\vec{Q}$
at the next time step is not ${\bf G} [ \vec{P} ]$, but is instead one
of the allowed lattice points in the finite-population state space. Since the
variance around the expected distribution ${\bf G}[\vec{P}]$ is
proportional to $1/M$, $\vec{Q}$ tends to be one of the lattice points
close to ${\bf G} [ \vec{P} ]$.

\subsection{Epochal Dynamics}

For finite populations, the expected change $\langle d\vec{P} \rangle$
in the fitness distribution over one generation is given by:
\begin{equation}
\langle d\vec{P}\rangle = {\bf G} [ \vec{P} ] -\vec{P}.
\end{equation}
Assuming that some component $\langle d P_i \rangle$ is much smaller
than $1/M$, the actual change in component $P_i$ is likely to be $d
P_i = 0$ for a long succession of generations. That is, if the size of
the ``flow'' $\langle dP_i\rangle$ in some direction $i$ is much smaller
than the lattice spacing ($1/M$) for the finite population, we expect
the fitness distribution to not change in direction (fitness) $i$.

In Refs. \cite{Nimw97a} and \cite{Nimw97b} we showed this is the
mechanism by which finite populations cause epochal dynamics. For the
Royal Staircase fitness functions, we have that whenever fitness $n$
is the highest in the population, such that $P_i = 0$ for all $i > n$,
the rate at which higher fitness strings are discovered is very small:
$\langle d P_i \rangle \ll 1/M$ for all $i > n$ and population size $M$
not too large. A period of stasis (an evolutionary epoch) thus
corresponds to the time the population spends before it discovers a
higher fitness string. More formally, each epoch $n$ corresponds to
the population being restricted to a region in the $n$-dimensional
lower-fitness subspace consisting of fitnesses $1$ to $n$ of the
macroscopic state space. Stasis occurs because the flow out of this
subspace is much smaller than the finite-population induced lattice
spacing.

As the experimental runs of Fig. \ref{fig_runs} illustrated, each
epoch in the average fitness is associated with a (typically) constant
value of the best fitness in the population. More detailed experiments
reveal that not only is $\langle f \rangle$ constant on average during
the epochs, in fact the entire fitness distribution $\vec{P}$
fluctuates in an approximately Gaussian way around some constant
fitness distribution $\vec{P}^n$ during the epoch $n$---the
generations when $n$ is the highest fitness in the population.

As was shown in Ref. \cite{Nimw97b}, each epoch fitness distribution
$\vec{P}^n$ is the unique fixed point of the operator ${\bf G}$
restricted to the $n$-dimensional subspace of strings with $1 \leq f
\leq n$. That is, if ${\bf G}^n$ is the projection of the operator
${\bf G}$ onto the $n$-dimensional subspace of fitnesses from $1$ up
to $n$, then we have:
\begin{equation}
{\bf G}^n [ \vec{P}^n ] = \vec{P}^n ~.
\end{equation}
By Eq. (\ref{AvFitness}), then, the average fitness $f_n$ in epoch
$n$ is given by:
\begin{equation}
f_n = \sum_{j=1}^{n} j P^n_j.
\end{equation}
Thus, the fitness distributions $\vec{P}^n$ during epoch $n$ are
obtained by finding the fixed point of ${\bf G}$ restricted to the
first $n$ dimensions of the fitness distribution space. We will pursue
this further in the next section.

To summarize at this point, the statistical dynamics analysis is
tantamount to the following qualitative picture. The global dynamics
can be viewed as an incremental discovery of successively more
(macroscopic) dimensions of the fitness distribution space. Initially, only strings
of low fitness are present in the initial population. The population
stabilizes on the epoch fitness distribution $\vec{P}^n$ corresponding
to the best fitness $n$ in the initial population. The fitness
distribution fluctuates around the $n$-dimensional vector $\vec{P}^n$
until a string of fitness $n+1$ is discovered and spreads through the
population. The population then settles into $(n+1)$-dimensional
fitness distribution $\vec{P}^{n+1}$ until a string of fitness $n+2$
is discovered, and so on, until the global optimum at fitness $N+1$ is
found. In this way, the global dynamics can be seen as stochastically
hopping between the different epoch distributions $\vec{P}^n$, unfolding
a new macroscopic dimension of the fitness distribution space each
time a higher fitness string is discovered.

Whenever mutation creates a string of fitness $n+1$, this string may
either disappear before it spreads, seen as the transient jumps in
best fitness in Fig. \ref{fig_runs}, or it may spread, leading the
population to fitness distribution $\vec{P}^{n+1}$. We call the latter
process an {\em innovation}. Through an innovation, a new
(macroscopic) dimension of fitness distribution space becomes stable.
Fig. \ref{fig_runs} also showed that it is possible for the
population to fall from epoch $n$ (say) down to epoch $n-1$. This
happens when, due to fluctuations, all individuals of fitness $n$ are
lost from the population. We refer to this as a {\em destabilization}
of epoch $n$. Through a destabilization, a dimension can, so to
speak, collapse. For some parameter settings, such as shown in Figs.
\ref{fig_runs}(a) and \ref{fig_runs}(c), this is very rare. In these
cases, the time for the GA to reach the global optimum is mainly
determined by the time it takes to discover strings of fitness $n+1$
in each epoch $n$. For other parameter settings, however, such as in
Figs. \ref{fig_runs}(b) and \ref{fig_runs}(d), the destabilizations
play an important role in how the GA reaches the global optimum. In
these regimes, destabilization must be taken into account in
calculating search times. This is especially important in the current
setting since, as we will show, the optimized GA often operates in
this type of marginally stable parameter regime.

\section{Quasispecies Distributions and Epoch Fitness Levels}
\label{an_appr}

During epoch $n$ the quasispecies fitness distribution $\vec{P}^n$ is
given by a fixed point of the operator ${\bf G}^n$. To obtain this
fixed point we linearize the generation operator by taking out the
factor $\langle f \rangle$, thereby defining a new operator ${\bf
  \tilde{G}}^n$ via:
\begin{equation}
{\bf G}^n = \frac{1}{\langle f \rangle} {\bf \tilde{G}}^n,
\end{equation}
where $\langle f \rangle$ is the average fitness of the fitness
distribution that ${\bf G}^n$ acts upon; see App. \ref{SelectionOperator}.
The operator ${\bf \tilde{G}}^n$ is just an ordinary (linear) matrix
operator and the quasispecies fitness distribution $\vec{P}^n$ is
nothing other than the principal eigenvector of this matrix
(normalized in probability).
Conveniently, one can show that the principal eigenvalue $f_n$ of
${\bf \tilde{G}}^n$ is also the average fitness of the quasispecies
distribution. In this way, obtaining the quasispecies distribution
$\vec{P}^n$ reduces to calculating the principal eigenvector of the
matrix ${\bf \tilde{G}}^n$. Again, the reader is referred to Ref.
\cite{Nimw97b}.

The matrices ${\bf \tilde{G}}^n$ are generally of modest size: i.e.,
their dimension is smaller than the number of blocks $N$ and
substantially smaller than the dimension of genotype space. Due to
this we can easily obtain numerical solutions for the epoch fitnesses
$f_n$ and the epoch quasispecies distributions $\vec{P}^n$. For a
clearer understanding of the functional dependence of the epoch
fitness distributions on the GA's parameters, however, App.
\ref{EpochFitnessQuasispecies} recounts analytical approximations to
the epoch fitness levels $f_n$ and quasispecies distributions
$\vec{P}^n$ developed in Ref. \cite{Nimw98a}.

The result is that the average fitness $f_n$ in epoch $n$, which is
given by the largest eigenvalue, is equal to the largest diagonal
component of the analytical approximation to ${\bf \tilde{G}}^n$
derived in App. \ref{EpochFitnessQuasispecies}. That is,
\begin{equation}
f_n = n(1-q)^{(n-1)K} ~.
\end{equation}
The epoch quasispecies is given by:
\begin{equation}
P^n_i = \frac{(1-\lambda) n \lambda^{n-1-i}}{n \lambda^{n-1-i} - i}
\prod_{j=1}^{i-1} \frac{n \lambda^{n-j} - j}{n \lambda^{n-1-j} -j} ~,
\label{EpochQuasiSpecies}
\end{equation}
where $\lambda = (1-q)^K$ is the probability that a block will undergo
no mutations. For the following, we are actually interested in
the most-fit quasispecies component $P^n_n$ in epoch $n$. For this
component, Eq. (\ref{EpochQuasiSpecies}) reduces to
\begin{equation}
P^n_n =
\lambda^{n-1} \prod_{j=1}^{n-1} \frac{f_n- f_j}{f_n - \lambda f_j} ~,
\label{text_pnn_expression}
\end{equation}
where we have expressed the result in terms of of the epoch fitness
levels $f_j$.

\section{Mutation Rate Optimization}
\label{sec_mut_rate_opt}

In the previous sections we argued that the GA's behavior can be viewed
as (occasionally) stochastically hopping from epoch to epoch---when the
search discovers a string with increased fitness that spreads
in the population. Assuming the total time
to reach this global optimum is dominated by the time the GA spends in
the epochs, Ref. \cite{Nimw98a} developed a way to tune the mutation
rate $q$ such that the time the GA spends in an epoch is minimized.
We briefly review this here before moving on to the more general theory
that includes population-size effects and epoch destabilization.

Optimizing the mutation rate amounts to finding a balance between
two opposing effects of varying mutation rate. On the one hand,
when the mutation rate is increased, the average number of mutations in
the unaligned blocks goes up thereby increasing the probability of
creating a new aligned block. On the other hand, due to the increased
number of deleterious mutations, the equilibrium proportions $P^n_n$
of individuals in the highest fitness class during each epoch $n$
decreases.

In Ref. \cite{Nimw98a} we derived an expression for the
probability $C_{n+1}$ to create, over one generation in epoch $n$, a
string of fitness $n+1$ that will stabilize by spreading through the
population. This is given by
\begin{equation}
C_{n+1} = M P^n_n P_a \pi_n(\lambda) ~,
\label{corrected_opt_approx}
\end{equation}
where $P_a = (1-\lambda)/(2^K-1)$ is the probability of aligning a
block (App. \ref{MutationOperator}) and $\pi_n(\lambda)$ is the
probability that a string of fitness $n+1$ will spread, as opposed
to being lost through a fluctuation or a deleterious mutation. This
latter probability largely depends on the relative average fitness
difference of epoch $n+1$ over epoch $n$. Denoting this difference as
\begin{equation}
\gamma_n = \frac{f_{n+1}-f_n}{f_n} = \left( 1 +
  \frac{1}{n}\right) \lambda -1,
\label{gamma_eq}
\end{equation}
and using a diffusion equation approximation (see Ref. \cite{Nimw97b}),
we found:
\begin{equation}
\pi_n(\lambda) = \frac{1 - \left( 1- \frac{1}{M}\right)^{2 M
    \gamma_n+1}}{1-\left(1- P^{n+1}_{n+1}\right)^{2 M \gamma_n+1}}.
\label{ProbToStabilize}
\end{equation}
If $P^{n+1}_{n+1} \gg 1/M$, this reduces to a population-size
independent estimate of the spreading probability
\begin{equation}
\pi_n \approx 1 - e^{-2 \gamma_n}.
\label{pop_indep_ProbToStabilize}
\end{equation}

If one allows for changing the mutation rate between epochs, one would
minimize the time spent in each epoch by maximizing $C_{n+1}$. Note
that $C_{n+1}$ depends on $q$ only through $\lambda$. The
optimal mutation rate in each epoch $n$ is determined by estimating
the optimal value $\lambda_o$ of $\lambda$ for each $n$. Although the
optimal $\lambda_o$ can be determined as the solution of an algebraic
equation, we found in Ref. \cite{Nimw98a} that it is well approximated by
\begin{equation}
\label{approx_lam_opt2}
\lambda_o(n) \approx 1 - \frac{1}{3 n^{1.175}} ~.
\end{equation}
For large $n$ this gave the optimal mutation rate as
\begin{equation}
\label{optimal_mutation}
q_o \approx \frac{1}{3 K n^{1.175}}~, n \gg 1 ~.
\end{equation}
Thus, the optimal mutation rate drops as a power-law in both $n$ and
$K$. This implies that if one is allowed to adapt the mutation rate
during the run, the mutation rate should decrease as a GA run
progresses so that the search will find the global optimum as quickly
as possible. We pursue this idea more precisely elsewhere by
considering an adaptive mutation rate scheme for a GA.

We now turn to the simpler problem of optimizing mutation rate for the
case of a {\em constant} mutation rate throughout a GA run. In Ref.
\cite{Nimw98a} we used Eq. (\ref{corrected_opt_approx}) to estimate the
total number $E$ of fitness function evaluations the GA uses on
average before an optimal string of fitness $N+1$ is found. As a first
approximation, we assumed that the GA visits all epochs, that the time
spent in innovations between them is negligible, and that epochs are
{\em always} stable. By epoch stability we that mean it is highly unlikely
that strings with the current highest fitness will disappear from the
population through a fluctuation, once such strings have spread.
These assumptions appear to hold for the parameters of Figs.
\ref{fig_runs}(a) and \ref{fig_runs}(c). They may hold even for the
parameters of Fig. \ref{fig_runs}(b), but they most likely do not for
Fig. \ref{fig_runs}(d). For the parameters of Fig. \ref{fig_runs}(d),
we see that the later epochs ($n = 9$, and $10$) easily destabilize
a number of times before the global optimum is found. Although we will
develop a generalization that addresses this more complicated behavior
in the next sections, it is useful to work through the optimization
of mutation rate first.

The average number $T_n$ of generations that the population spends in
epoch $n$ is simply $1/C_{n+1}$, the inverse of the probability that
a string of fitness $n+1$ will be discovered and spread through the
population. For a population of size $M$, the number of fitness
function evaluations per generation is $M$, so that the total number
$E_n$ of fitness function evaluations in epoch $n$ is given by $M T_n$.
More explicitly, we have:
\begin{equation}
E_n = \left( P^n_n P_a \pi_n \right)^{-1}.
\label{EnApprox}
\end{equation}
That is, the total number of fitness function evaluations in each
epoch is independent of the population size $M$. This is due to two
facts, given our approximations. First, the epoch lengths, measured in
generations, are inversely proportional to $M$, while the number of
fitness function evaluations per generation is $M$. Second, since for
stable epochs $P^n_n \gg 1/M$, the probability $\pi_n$ is also
independent of population size $M$; recall Eq.
(\ref{pop_indep_ProbToStabilize}).

The total number of fitness function evaluations $E(\lambda)$ to reach
the global optimum is simply given by substituting into
Eq. (\ref{EnApprox}) our analytical expressions for $P^n_n$ and
$\pi_n$, Eqs. (\ref{text_pnn_expression}) and
(\ref{pop_indep_ProbToStabilize}), and then summing $E_n(\lambda)$
over all epochs $n$ from $1$ to $N$. We found that:
\begin{equation}
\label{tot_lam}
E(\lambda)=\sum_{n=1}^{N} \frac{1}{P_a \pi_n(\lambda)} \prod_{i=1}^{n-1}
    \frac{n \lambda^{n-i-1} -i}{n \lambda^{n-i} -i} ~.
\end{equation}
Note that in the above equation $\pi_N = 1$ by definition because the
algorithm terminates as soon as a string of fitness $N+1$ is found.
That is, strings of fitness $N+1$ need not spread through the
population. The optimal mutation rate for an entire run is obtained by
minimizing Eq. (\ref{tot_lam}) with respect to $\lambda$.
\begin{figure}[htbp]
\centerline{\epsfig{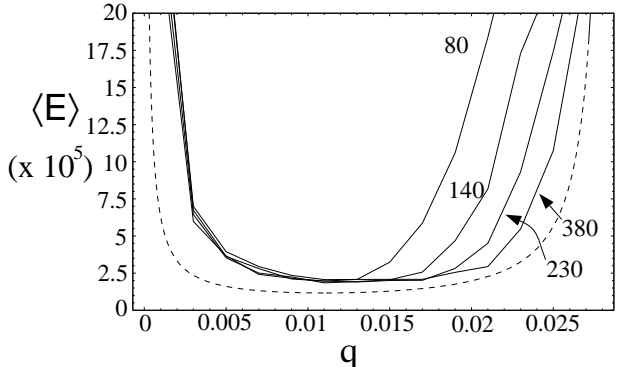}}
\caption{Average total number $\langle E \rangle$ of fitness function evaluations
  as a function of mutation rate $q$, from the theory (dashed), Eq.
  (\ref{tot_lam}), and from experimental estimates (solid). The fitness
  function parameter settings are $N=4$ blocks of length $K=10$ bits.
  The mutation rate runs from $q=0.001$ to $q=0.028$. Experimental
  data points are estimates over $250$ runs. The experimental curves
  show four different population sizes; $M=80$, $M=140$, $M=230$, and
  $M=380$.}
\label{mut_dep_fig}
\end{figure}

Figure \ref{mut_dep_fig} shows for $N=4$ blocks of length $K=10$ bits
the dependence of the average total number $E(q)$ of fitness function
evaluations on the mutation rate $q$. The dashed line is the theoretical
prediction of Eq. (\ref{tot_lam}); while the solid lines show the
experimentally estimated values of $\langle E \rangle$ for four
different population sizes. Each experimental data point is an estimate
obtained from $250$ GA runs. Figure \ref{mut_dep_fig} illustrates in a
compact form the findings of Ref. \cite{Nimw98a}, which can be
summarized as follows. 
\begin{enumerate}
\item{At fixed population size $M$, there is a smooth cost function
    $E(q)$ as a function of mutation rate $q$. It has a {\em single}
    and {\em shallow} minimum $q_o$, which is accurately predicted by
        the theory.}
\item{The curve $E(q)$ is everywhere concave.}
\item{The theory slightly underestimates the experimentally obtained
    $\langle E\rangle$.}
\item{The optimal mutation rate $q_o$ roughly occurs in the regime
    where the highest epochs are marginally stable; see Fig.
    \ref{fig_runs}.}
\item{For mutation rates lower than $q_o$ the experimentally estimated
    total number of fitness function evaluations $\langle E\rangle$
    grows steadily and becomes almost independent of the population
    size $M$. (This is where the experimental curves in Fig.
        \ref{mut_dep_fig} overlap.) For mutation rates larger than $q_o$
        the total number of fitness function evaluations does depend on $M$,
        which is not explained by the theory of Ref. \cite{Nimw98a}.}
\item{There is mutational error threshold in $q$ that bounds the
        upper limit in $q$ of the GA's efficient search regime. Above
        the threshold, which is population-size independent,
        suboptimal strings of fitness $N$ cannot stabilize in the
        population, even for very large population sizes. This error
        threshold is also correctly predicted by the theory. It occurs
        around $q_c = 0.028$ for $N=4$ and $K=10$.}
\end{enumerate}

\section{Epoch Destabilization: Population-Size Dependence}
\label{destab_sec}

We now extend the above analysis to account for $E$'s dependence on
population size. This not only improves the parameter-optimization
theory, but also leads us to consider a number of issues and mechanisms
that shed additional light on how GAs work near their optimal parameter
settings. Since it appears that optimal parameter settings often lead
the GA to run in a behavioral regime were the population dynamics is
marginally stable in the higher epochs, we consider how
destabilization dynamics affects the time to discover the global
optimum.

We saw in Figs. \ref{fig_runs}(b) and \ref{fig_runs}(d) that, around
the optimal parameter settings, the best fitness in the population can
show intermittent behavior. Apparently, fluctuations sometimes cause
an epoch's current best strings (of fitness $n$) in the population to
disappear. The best fitness then drops to $n-1$. Often, strings of
fitness $n$ are rediscovered later on. Qualitatively, what happens
during these destabilizations is that, since the proportion
$P^n_n$ of individuals in the highest fitness class decreases for
increasing $n$ and $q$ (Eq. (\ref{text_pnn_expression})), for small
population sizes the absolute number of individuals in the highest
fitness class approaches a single string; i.e., $M P^n_n \approx 1$ in
higher epochs. When this happens, it is likely that all individuals of
fitness $n$ are lost through a deleterious fluctuation and the
population falls back onto epoch $n-1$. Somewhat more precisely,
whenever the standard deviation of fluctuations in the proportion
$P_n$ of individuals with fitness $n$ becomes as small as their
equilibrium proportion $P^n_n$, destabilizations start to become a
common occurrence. 

Since the probability of a destabilization is sensitive to the
population size $M$, this dynamical effect introduces population-size
dependence in the average total number $\langle E \rangle$ of fitness
function evaluations.

As we just noted, the theory for $E(q)$ used in Ref. \cite{Nimw98a}
assumed that all epochs are stable, leading to a population-size
independent theory. However, as is clear from Fig. \ref{fig_runs}(d),
one should take into account the (population-size dependent)
probability of epoch $n$ destabilizing several times to epoch $n-1$
before the population moves to epoch $n+1$. For example, if during
epoch $n$ the population is $3$ times as likely to destabilize to epoch
$n-1$ compared to innovating to epoch $n+1$, then we expect epoch $n$
to disappear three times before moving to epoch $n+1$. Assuming that
epoch $n-1$ is stable, this increases the number of generations spent
in epoch $n$ by roughly three times the average number of generations
spent in epoch $n-1$.

To make these ideas precise we introduce a Markov chain model to
describe the ``hopping'' up and down between the epochs. The Markov
chain has $N+1$ states, each representing an epoch. In every generation
there are probabilities $p^+_{n}$ to innovate from epoch $n$ to epoch
$n+1$ and $p^-_{n}$ to destabilize, falling from epoch $n$ to epoch
$n-1$. The globally optimum state $N+1$ is an absorbing state. Starting
from epoch $1$ we calculate the expected number $T$ of generations for
the population to reach the absorbing state for the first time. 

The innovation probabilities $p^+_n$ are just given by the $C_{n+1}$
of Eq. (\ref{corrected_opt_approx}):
\begin{equation}
p^+_n = C_{n+1} = \frac{M}{E_n},
\end{equation}
where $E_n$ is given by the approximation of Eq. (\ref{EnApprox}).
Note that when $M P^n_n$ approaches $1$ the spreading probability
$\pi_n$, as given by Eq. (\ref{ProbToStabilize}), becomes
population-size dependent as well, and we use Eq. (\ref{ProbToStabilize})
rather than Eq. (\ref{pop_indep_ProbToStabilize}). To obtain the
destabilization probabilities $p^-_n$ we assume that in each generation
the population has an equal and independent probability to destabilize
to epoch $n-1$. This probability is given by the inverse of the average
time until a destabilization occurs.

In Ref. \cite{Nimw97b} we studied the destabilization mechanism using a
diffusion equation method. We derived an analytical approximation for
the average number of generations $D_n$ until epoch $n$ destabilizes
and falls back onto epoch $n-1$. The result is:
\begin{eqnarray}
D_n & = & \frac{M P^n_n}{1-P^n_n} \nonumber \\
        & + & \frac{\pi}{2 \mu_n} {\rm erfi} \left[
  \sqrt{\frac{M \mu_n P^n_n}{1 - P^n_n}} \right] {\rm erf} \left[ \sqrt{\frac{
  M \mu_n (1-P^n_n)}{P^n_n}}\right],
\label{destab_equation}
\end{eqnarray}
where ${\rm erf}(x)$ is the error function and
${\rm erfi}(x) = {\rm erf}(i x)/i$ is the imaginary error function.
  
In Ref. \cite{Nimw97b} we pointed out the connection between the above
formula and error thresholds in the theory of molecular evolution.
Generally, error thresholds denote the boundary in parameter space
between a regime where a certain high fitness string, or an equivalence
class of high fitness strings, is stable in the population, and a regime
where it is unstable. In the case of a single high fitness ``master
sequence'' one speaks of a genotypic error threshold; see Refs.
\cite{Alves&Fontanari98,Eigen&McCasKill&Schuster89,Nowak&Schuster89,Swetina&Schuster82}.
In the case of an equivalence class of high fitness strings, one speaks
of a {\em phenotypic} error threshold; see Refs.
\cite{Huynen&Stadler&Fontana,Reidys98b}.

A sharply defined error threshold generally only occurs in the limit of
infinite populations and infinite string length \cite{Leuthausser87},
but extensions to finite population cases have been studied in Refs.
\cite{Alves&Fontanari98,Nowak&Schuster89,Reidys98b}. In Ref.
\cite{Reidys98b}, for example, the occurrence of a finite-population
phenotypic error threshold was defined by the equality of the standard
deviation and the mean of the number of individuals of the highest
fitness class. This definition is in accord with Eq.
(\ref{destab_equation}): the argument of ${\rm erfi(x)}$,
$\sqrt{M \mu_n P^n_n / (1 - P^n_n)}$, is exactly the ratio between the
mean proportion $P^n_n$ and standard deviation of the number of
individuals with fitness $n$, as derived in Ref. \cite{Nimw97b}. The
function ${\rm erfi}(x)$ is a very rapidly growing function of its
argument: ${\rm erfi}(x) \approx \exp( x^2)/x$ for $x$ larger than $1$.
Therefore, $\sqrt{M \mu_n P^n_n / (1 - P^n_n)}$ being either smaller
(larger) than $1$ is a reasonable criterion for the instability
(stability) of an epoch. Of course, this is simply a way of summarizing
the more detailed information contained in Eq. (\ref{destab_equation}).
  
The constant $\mu_n$ in Eq. (\ref{destab_equation}) is the average
decay rate of fluctuations in the number of individuals in the highest
fitness class around its equilibrium value $P^n_n$. The value of
$\mu_n$ for epoch $n$ can be calculated in terms of the relative sizes
of the fluctuations in the directions of all lower-lying epochs. This
calculation was performed explicitly in Ref. \cite{Nimw97b}. Formally,
one needs to rotate the covariance matrix of sampling fluctuations
during epoch $n$ to the basis of epoch eigenvectors $\vec{P}^i$.  The
covariance matrix of sampling fluctuations during epoch $n$ is
approximately given by:
\begin{equation}
\langle d P_i d P_j \rangle = \frac{P^n_i (\delta_{ij}-P^n_j)}{M}.
\end{equation}
Defining the matrix ${\bf R}$ such that
its columns contain the epoch distributions $\vec{P}^j$:
\begin{equation}
{\bf R}_{ij} = P^j_i,
\end{equation}
we can rotate the covariance matrix to the basis of epoch vectors by
using the inverse of ${\bf R}$.
The vector $\vec{B}$ contains the diagonal components of this rotated
covariance matrix:
\begin{equation}
B_i = \frac{1}{M} \sum_{k,m=1}^{n-1} {\bf R}^{-1}_{ik} {\bf R}^{-1}_{im} P^n_k
\left( \delta_{km} -\vec{P}^n_m \right).
\end{equation}
The components $B_i$ are proportional to the amplitude of fluctuations
in the direction of epoch $i$ during epoch $n$. The decay rate of
fluctuations in the direction of epoch $i$ is given by $(f_n -f_i)/f_n$.
The decay rate $\mu_n$ is then simply given by the average decay rates
of fluctuations in the directions of the lower lying epochs, weighted
by the sampling fluctuations vector $\vec{B}$. That is,
\begin{equation}
\mu_n = \frac{\sum_{i=1}^{n-1} (f_n-f_i) B_i}{f_n \sum_{i=1}^{n-1}
  B_i}.
\end{equation}
Generally, $\mu_n$ decreases monotonically as a function of $n$
since fluctuations in the proportion $P^n_n$ of individuals in the
highest fitness class $n$ decay more slowly for higher epochs. 

Thus, we have for the destabilization probabilities:
\begin{equation}
p^-_n = \frac{1}{D_n}.
\end{equation}
Finally, note that the probability to remain in epoch $n$ is
$1-p^+_n -p^-_n$.

With these expressions for the transition probabilities of the
Markov chain, it is now straightforward to calculate the
average number $T$ of generations before the GA discovers the global
optimum for the first time; see for instance Sec. 7.4 of Ref.
\cite{Gardiner85}. The solution is:
\begin{equation}
\label{T_sol}
T = \sum_{n=1}^{N} \phi_n \sum_{k=1}^{n} \frac{1}{p^+_k \phi_k},
\end{equation}
where $\phi_n$ is defined as:
\begin{equation}
\phi_n = \prod_{k=2}^n \frac{p^-_k}{p^+_k}, ~ n \geq 2 ~,
\end{equation}
and
\begin{equation}
\phi_1 = 1.
\end{equation}
Since Eq. (\ref{T_sol}) gives the average number $T$ of generations,
the average number of fitness function evaluations
$E(q,M)$ is given by:
\begin{eqnarray}
\label{M_dep_E_sol}
E(q,M) & = & M T \nonumber \\
           & = & E_N + E_{N-1} \left(1 + \frac{E_N}{M D_N}\right)
           \nonumber \\
       & + & E_{N-2}\left( 1 + \frac{E_{N-1}}{M D_{N-1}}
       \left( 1 + \frac{E_N}{M D_N}\right)\right) \nonumber \\
           & + & \ldots ~,
\end{eqnarray}
where $E_n$ is given by Eq. (\ref{EnApprox}) and where the last equality
is obtained by rewriting the sums in Eq. (\ref{T_sol}). It is easy to
see that as epochs become arbitrarily stable ($D_n \rightarrow \infty$)
this solution reduces to Eq. (\ref{tot_lam}). 

\section{Theory versus Experiment}

\begin{figure}[htbp]
\centerline{\epsfig{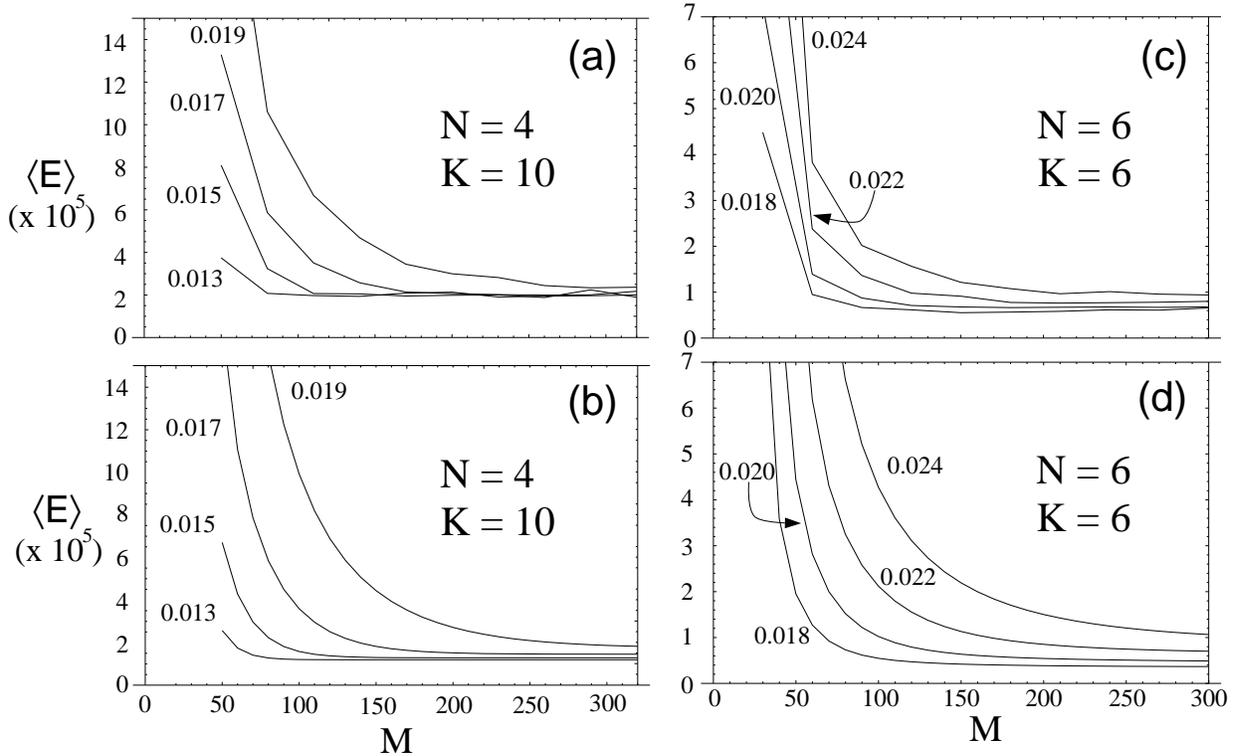}}
\caption{Average total number $\langle E \rangle$ of fitness function
  evaluations as a function of the population size $M$ for two different
  fitness function parameters and four mutation rates each, both
  experimentally ((a) and (c), top row) and theoretically ((b) and (d),
  bottom row). In each figure each solid line
  gives $E(M)$ for a different mutation rate. Each experimental data
  point is an average over $250$ GA runs. Figures (a) and (b) have
  $N=4$ blocks of length $K=10$. The upper figure (a) shows the
  experimentally estimated $E(M)$ as a function of $M$ for the mutation
  rates $q \in \{0.013,0.015,0.017,0.019\}$. The lower figure (b) shows
  the theoretical results, as given by Eq. (\ref{M_dep_E_sol}), for
  the same parameter settings. In both, the population size ranges from
  $M=50$ to $M=320$ on the horizontal axis. Figures (c) and (d) have
  $N=6$ blocks of length $K=6$. Figure (c) shows the experimental
  averages and figure (d) the theoretical predictions for the same
  parameter settings. The population sizes on the horizontal axis run
  from $M=30$ to $M=300$. The mutation rates shown in (c) and (d) are
  $q \in \{0.018,0.02,0.022,0.024\}$.}
\label{pop_dep_mosaic}
\end{figure}

We can now compare this population-size dependent approximation, Eq.
(\ref{M_dep_E_sol}), with the experimentally measured dependence on
$M$ of the average total number $\langle E \rangle$ of fitness
function evaluations. Figure \ref{pop_dep_mosaic} shows the dependence
of $\langle E \rangle$ on the population size $M$ for two different
parameter settings of $N$ and $K$ and for a set of mutation rates $q$.

The upper figures, Figs. \ref{pop_dep_mosaic}(a) and
\ref{pop_dep_mosaic}(c), give the dependence of the experimentally
estimated $\langle E \rangle$ on the population size $M$. The lower
figures, Figs. \ref{pop_dep_mosaic}(b) and \ref{pop_dep_mosaic}(d), give
the theoretical predictions from Eq. (\ref{M_dep_E_sol}). The upper left
figure, Fig. \ref{pop_dep_mosaic}(a), shows $\langle E \rangle$ as a
function of $M$ for $N=4$ blocks of length $K=10$ for four different
mutation rates: $q \in \{0.013,0.015,0.017,0.019\}$. The population
size ranges from $M=50$ to $M=320$. The total number of fitness function
evaluations on the vertical axis ranges from $\langle E \rangle = 0$
to $\langle E \rangle = 15 \times 10^5$. Each data point was obtained as
an average over $250$ GA runs. Figure \ref{pop_dep_mosaic}(b) shows the
theoretical predictions for the same parameter settings. Figure
\ref{pop_dep_mosaic}(c) gives the experimental estimates for $N=6$
blocks of length $K=6$, over the range $M=30$ to $M=300$, for four
mutation rates: $q \in \{0.018,0.02,0.022,0.024\}$. The total number
of fitness function evaluations on the vertical axis range from
$\langle E \rangle = 0$ to $\langle E \rangle = 7 \times 10^5$. Figure
\ref{pop_dep_mosaic}(d) shows the theoretical predictions for the same
range of $M$ and the same four mutation rates.

We see that as the population size becomes ``too small''
destabilizations make the total number of fitness function evaluations
increase rapidly. The higher the mutation rate, the higher the
population size at which the sharp increase in $\langle E \rangle$
occurs. These qualitative effects are captured accurately by the
theoretical predictions from Eq. (\ref{M_dep_E_sol}). Although our
analysis involves several approximations (e.g. as in the calculations of
$D_n$), the theory does quantitatively capture the population-size
dependence well, both with respect to the predicted absolute number of
fitness function evaluations and the shape of the curves as a function
of $M$ for the different mutation rates. From Figs.
\ref{pop_dep_mosaic}(c) and \ref{pop_dep_mosaic}(d) it seems that the
theory overestimates the growth of $\langle E \rangle$ for the larger
mutation rates as the population size decreases. Still, the theory
correctly captures the sharp increase of $\langle E \rangle$ around a
population size of $M=50$.

As the population size increases beyond approximately $M=200$, we find
experimentally that the average total number of fitness function
evaluations $\langle E \rangle$ starts rising slowly as a function of
$M$. This effect is not captured by our analysis. It is also barely
discernible in Figs. \ref{pop_dep_mosaic}(a) and \ref{pop_dep_mosaic}(c).
We believe that the slow increase of $\langle E \rangle$ for large
population sizes derives from two sources.

First, by the maximum entropy assumption, our theory assumes that all
individuals in the highest fitness class are genetically
{\em independent}, apart from the sharing of their aligned blocks.
This is not true in general. The sampling of the selection process
introduces genetic correlations in the individuals of the highest
fitness class. Under our assumption of independence, doubling the
population size from $M$ to $2 M$ should reduce the number of
generations to find the global optimum by an equal factor of $2$,
making $\langle E \rangle$ independent of $M$.
In reality, since the strings in the highest fitness class are
correlated, doubling the population size from $M$ to $2 M$ will reduce
the total number of generations by a factor somewhat {\em less} than
two, thereby increasing $\langle E \rangle$ slightly. Unfortunately,
this effect is very hard to address
quantitatively.

\end{multicols}

\begin{multicols}{2}

The second reason for the increase of $E$ with increasing population
size comes from the time the population spends in the short
innovations between the different epochs. Up to now, we have
neglected these innovation periods. Generally, they only contribute
marginally to $E$. In Ref. \cite{Nimw97b} we
calculated the approximate number of generations $g_n$ that the
population spends in the innovation from epoch $n$ to epoch $n+1$ and
found that:
\begin{equation}
g_n = \frac{2 + \gamma_n}{\gamma_n} \log\left[ M \right],
\end{equation}
where $\gamma_n$ is the fitness differential given by Eq.
(\ref{gamma_eq}). The GA thus expends a total of
\begin{equation}
{\bf I} = M \log \left[ M \right] \sum_{n=1}^{N-1} \frac{2 +
  \gamma_n}{\gamma_n},
\label{innov_cor}
\end{equation}
fitness function evaluations in the innovations. Notice that this
number grows as $M \log \left[ M \right]$. Since the terms in the
above sum are generally much smaller than $E_n$, the contribution of
$\bf I$ only leads to a slow increase in $\langle E \rangle$ as $M$
increases.

\section{Search-Effort Surface and Generalized Error Thresholds}

\begin{figure}[htbp]
\centerline{\epsfig{file=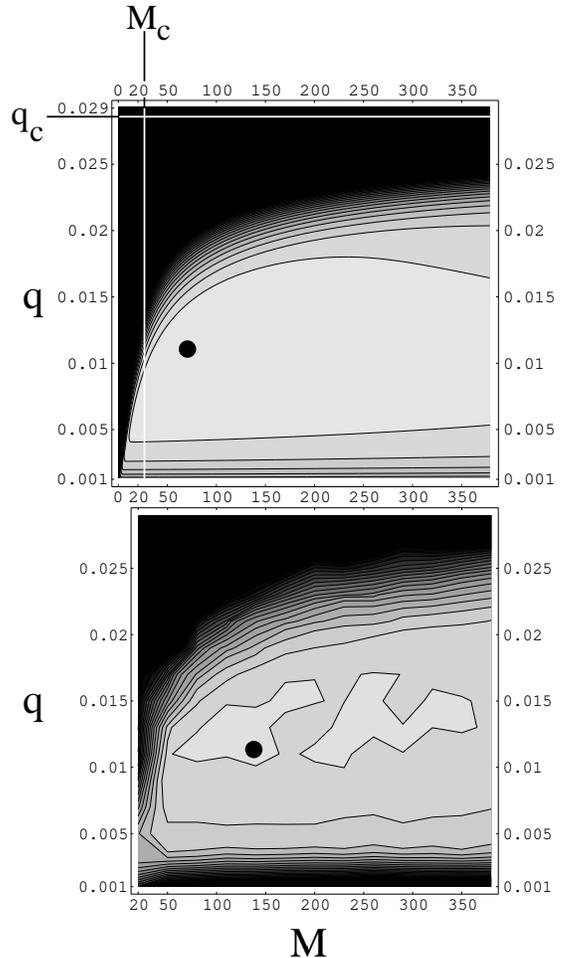,width=4.0in}}
\caption{Contour plots of the search-effort surface $E(q,M)$ of the average total number
  of fitness function evaluations for the theory (upper), Eqs.
  (\ref{M_dep_E_sol}) and (\ref{innov_cor}), and for experimental
  estimates (lower). The
  parameter settings are $N=4$ blocks of length $K=10$ bits. The
  population size $M$ runs from $M=1$ to $M=380$ on the horizontal
  axis on the upper plot and from $M=20$ to $M=380$ on the lower.  The
  mutation rate runs from $q=0.001$ to $q=0.029$ on the vertical.  The
  contours are plotted over the range $E(q,M) = 0$ to $E(q,M) = 2
  \times 10^6$ with a contour at each multiple of $10^5$. The
  experimental surface was interpolated from $195$ equally spaced data
  points, $13$ increments of $\Delta M = 30$ on the horizontal axis by
  $15$ increments of $\Delta q = 0.002$ on the vertical. The
  theoretical surface was interpolated over a grid using $\Delta M =
  1$ and $\Delta q = 0.00025$. The optimal theoretical parameter
  setting, $(q_o,M_o) = (0.011,60)$, and the optimal experimental
  parameter setting, $(q_o,M_o) = (0.011,140)$, are marked in their
  respective plots with a dot.}
\label{contours_fig}
\end{figure}

We summarize our theoretical and experimental findings for the entire
{\em search-effort surface} $E(q,M)$ of the average total number of
fitness function evaluations in Fig. \ref{contours_fig}.

The figure shows the average total number $E(q,M)$ of fitness function
evaluations for $N=4$ blocks of length $K=10$ bits; the same fitness
function as used in Figs. \ref{fig_runs}(c), \ref{mut_dep_fig},
\ref{pop_dep_mosaic}(a), and \ref{pop_dep_mosaic}(b). The top plot
shows the theoretical predictions, which now include the innovation
time correction from Eq. (\ref{innov_cor}); the bottom, the experimental
estimates. The horizontal axis ranges from a population size of $M=1$
($M=20$, experimental) to a population size of $M=380$ with steps of
$\Delta M = 1$ ($\Delta M = 30$, experimental). The vertical axis runs
from a mutation rate of $q=0.001$ to $q=0.029$ with steps of $\Delta q
= 0.00025$ in the theoretical plot and $\Delta q = 0.002$ in the
experimental. The experimental search-effort surface is thus an
interpolation between $195$ data points on an equally spaced lattice
of parameter settings. Each experimental data point is an average
over $250$ GA runs. The contours range from $E(q,M)= 0$ to $E(q,M) = 2
\times 10^6$ with each contour representing a multiple of $10^5$. Note
that the lowest values of $E$ lie between $10^5$ and $2 \times 10^5$.
Lighter gray scale corresponds to smaller values of $E(q,M)$.

The initial observations from Fig. \ref{contours_fig} were already
apparent from Fig. \ref{mut_dep_fig} and Fig. \ref{pop_dep_mosaic}.
First, the theory correctly predicts the relatively large region in
parameter space where the GA searches most efficiently. Second, the
theory correctly predicts the location of the optimal parameter
settings, indicated by a dot in the upper plot of Fig.
\ref{contours_fig}. The optimum occurs for somewhat higher population
size in the experiments, as indicated by the dot in the lower plot of
Fig. \ref{contours_fig}. Due to the large variance in $E$ from run to
run (recall Table \ref{FitEvalsTable}) and the rather small differences
in the experimental values of $\langle E\rangle$ near this regime,
however, it is hard to infer from the experimental data exactly where
the optimal population size occurs. Third, the theory
underestimates the absolute magnitude of $E(q,M)$ somewhat. Fourth, at
small mutation rates $E(q,M)$ increases more slowly for decreasing $q$
in the theoretical plot than in the experimental plot. Apart from
this, though, the plots illustrate the general shape of the
search-effort surface $E(q,M)$.

There is a relatively large area of parameter space around the optimal
setting $(q_o,M_o)$ for which the GA runs efficiently. Moving away from
this optimal setting horizontally (changing $M$) increases $E(q,M)$ only
slowly at first. For decreasing $M$ one reaches a ``wall'' relatively
quickly around $M=30$. For population sizes lower than $M=30$, the
higher epochs become so dynamically unstable that it is difficult for
the population to reach the global optimum string {\em at all}. In
contrast, moving in the opposite direction, increasing population
size, $E(q,M)$ increases slowly over a relatively
large range of $M$. Thus, choosing the population size too small is
far more deleterious than setting it too large.

Moving away from the optimal setting vertically (changing $q$) the
increase of $E(q,M)$ is also slow at first.
Eventually, as the plots make clear, increasing $q$ one reaches the
same ``wall'' as encountered in lowering $M$. This occurs at $q
\approx 0.02$ in Fig. \ref{contours_fig}. For larger mutation rates
the higher epochs become too unstable in this case as well, and the
population is barely able to reach the global optimum.

The wall in $(q,M)$-space is the two-dimensional analogue of a
phenomenon known as the {\em error threshold} in the theory of
molecular evolution. As pointed out in Sec. \ref{destab_sec}, in
our case error thresholds form the boundary between parameter regimes
where epochs are stable and unstable. Here, the {\em error boundary}
delimits a regime in parameter space where the optimum is discovered
relatively quickly from a regime, in the black upper left corners of the
plots, where the population essentially never finds the optimum. For
too high mutation rates or too low population sizes, selection is not
strong enough to maintain high fitness strings---in our case those
close to the global optimum---in the population against sampling
fluctuations and deleterious mutations. Strings of fitness $N$ will
not stabilize in the population but will almost always be immediately
lost, making the discovery of the global optimum string of fitness
$N+1$ extremely unlikely.

Note that the error boundary rolls over with increasing $M$ in the
upper left corner of the plots. It bends all the way over to the
right, eventually running horizontally, thereby determining a
population-size {\em independent} error threshold. For our parameter
settings this occurs around $q \approx 0.028$. Thus, beyond a critical
mutation rate of $q_c \approx 0.028$ the population almost never
discovers the global optimum, even for very large populations.

The value of this horizontal asymptote $q_c$ can be roughly approximated
by calculating for which mutation rate $q_c$ epoch $N$ has exactly the
same average fitness as epoch $N-1$; i.e. find $q_c$ such that $f_N
\approx f_{N-1}$. For those parameters, the population is under no
selective pressure to move from epoch $N-1$ to epoch $N$. Thus,
strings of fitness $N$ will generally not spread in the population.
Using our analytic approximations, we find that the critical mutation
rate $q_c$ is simply given by:
\begin{equation}
q_c = 1 - \sqrt[K]{\frac{N-1}{N}} ~.
\end{equation}
For the parameters of Fig. \ref{contours_fig} this gives $q_c =
0.0284$. This asymptote is indicated there by the horizontal line in
the top plot.

Similarly, below a critical population size $M_c$, it is also
practically impossible to reach the global optimum, even for low
mutation rates. This $M_c$ can also be roughly approximated by
calculating the population size for which the sampling noise is equal
to the fitness differential between the last two epochs. We find:
\begin{equation}
M_c = \left( \frac{N-1}{N \lambda - N + 1} \right)^2 ~.
\end{equation}
For the parameters of Fig. \ref{contours_fig} this gives
$M_c \approx 27$ around $q=0.011$. This threshold estimate is indicated
by the vertical line in Fig. \ref{contours_fig}.

Further, notice that for small mutation rates, at the bottom of each
plot in Fig. \ref{contours_fig}, the contours run almost
horizontally. That is, for small mutation rates relative to the
optimum mutation rate $q_o$, the total number of fitness function
evaluations $E(q,M)$ is insensitive to the population size $M$.
Decreasing the mutation rate too far below the optimum rate increases
$E(q,M)$ quite rapidly. According to our theoretical predictions it
increases roughly as $1/q$ with decreasing $q$. The experimental data
indicate that this is a slight underestimation. In fact, $E(q,M)$
appears to increase as $1/q^{\alpha}$ where the exponent $\alpha$ lies
somewhere between $1$ and $2$.

Globally, the theoretical analysis and empirical evidence indicate that
the search-effort surface $E(q,M)$ is everywhere concave. That is, for any two
points $(q_1,M_1)$ and $(q_2,M_2)$, the straight line connecting these
two points is everywhere above the surface $E(q,M)$. We believe that
this is always the case for mutation-only genetic algorithms with a
static fitness function that has a unique global optimum. This feature
is useful in the sense that a steepest descent algorithm on the level
of the GA {\em parameters} $q$ and $M$ will always lead to the unique
optimum $(q_o,M_o)$.

Finally, it is important to emphasize once more that there are large
run-to-run fluctuations in the total number of fitness evaluations to
reach the global optimum. (Recall Table \ref{FitEvalsTable}.)
Theoretically, each epoch has an exponentially distributed length since
there is an equal and independent innovation probability of leaving it
at each generation. The standard
deviation of an exponential distribution is equal to its mean. Since
the total time $E(q,M)$ is dominated by the last epochs, the total
time $E(q,M)$ has a standard deviation close to its mean.

One conclusion from this is that, if one is only going to use a GA for
a few runs on a specific problem, there is a large range in parameter
space for which the GA's performance is statistically equivalent. In
this sense, fluctuations lead to a large ``sweet spot'' of GA
parameters. On the other hand, these large fluctuations reflect the
fact that individual GA runs do not reliably discover the global
optimum within a fixed number of fitness function evaluations.

\section{Conclusions}

We derived explicit analytical approximations to the total number of
fitness function evaluations that a GA takes on average to discover
the global optimum as a function of both mutation rate and population
size. The class of fitness functions so analyzed describes a very
general subbasin-portal architecture in genotype space. We found that
the GA's dynamics consists of alternating periods of stasis (epochs)
in the fitness distribution of the population, with short bursts of
change (innovations) to higher average fitness. During the epochs the
most-fit individuals in the population diffuse over neutral networks
of isofitness strings until a portal to a network of higher fitness is
discovered. Then descendants of this higher fitness string spread
through the population.

The time to discover these portals depends both on the fraction of the
population that is located on the highest neutral net in equilibrium
and the speed at which these population members diffuse over the
network. Although increasing the mutation rate increases the
diffusion rate of individuals on the highest neutral network, it also
increases the rate of deleterious mutations that cause these members
``fall off'' the highest fitness network. The mutation rate is
optimized when these two effects are balanced so as to maximize the
total amount of explored volume on the neutral network per generation.
The optimal mutation rate, as given by Eq. (\ref{optimal_mutation}),
is dependent on the neutrality degree (the local branching rate) of
the highest fitness network and on the fitnesses of the lower lying
neutral networks onto which the mutants are likely to fall.

With respect to optimizing population size, we found that the optimal
population size occurs when the highest epochs are just barely stable.
That is, given the optimal mutation rate, the population size should
be tuned such that only a few individuals are located on the highest
fitness neutral network. The population size should be large enough
such that it is relatively unlikely that all the individuals disappear
through a deleterious fluctuation, but not much larger than that. In
particular, if the population is much larger, so that many individuals
are located on the highest fitness network, then the sampling dynamics
causes these individuals to correlate genetically. Due to this genetic
correlation, individuals on the highest fitness net do not independently
explore the neutral network. This leads, in turn, to a deterioration of
the search algorithm's efficiency. Therefore, the population size should
be as low as possible without completely destabilizing the last epochs.
Given this, one cannot help but wonder how general the association of
efficient search and marginal stability is.

It would appear that the GA wastes computational resources when
maintaining a population quasispecies that contains many suboptimal
fitness members; that is, those that are not likely to discover higher
fitness strings. This is precisely the reason that the GA performs so
much more poorly than a simple hill climbing algorithm on this
particular set of fitness functions, as shown in Ref.
\cite{MitchellEtAl93d}. The deleterious mutations together with the
nature of the selection mechanism drives up the fraction of lower
fitness individuals in the quasispecies. If we allowed ourselves to
tune the selection strength, we could have tuned selection so high
that only the most-fit individuals would ever be present. As we will
show elsewhere, this leads to markedly better performance, equal to or
even slightly exceeding that of hill climbing algorithms. Thus, the
GA's comparatively poor performance is the result of resources being
wasted on the presence of suboptimal fitness individuals.

In contrast, the reason that only the best individuals must be kept for
optimal search is a result of the fact that our set of fitness functions
has no local optima. If there are small fluctuations in fitness on the
neutral networks or if there is noise in the fitness function
evaluation, it might be beneficial to keep some of the lower fitness
individuals in the population. We will also pursue this elsewhere.

For now, it suffices to recall once more the typical dynamical
behavior of the GA population around the optimal parameter settings.
The GA searches most efficiently when population size and mutation
rate are set such that the epochs are marginally stable. That is, the
GA dynamics is as ``stochastic'' as possible without destabilizing the
current and later epochs. Strings of fitness $n$ are (only slightly)
preferentially reproduced over strings with fitness $n-1$, and the
population size is just large enough to protect these fitness $n$
strings from deleterious sampling fluctuations.

More precisely,
mutation rate, population size, and network neutralities set a lower
bound $\delta f$ of fitness differentials that can be ``noticed'' by
the selection mechanism. This idea is closely related to so called
``nearly neutral'' theories of molecular evolution of Refs.
\cite{Ohta73,Ohta&Gillespie96}. For optimal parameter settings, the
fitness differential $\delta f = n- (n-1) = 1$ is just barely detected
by selection. Imagine that during epoch $n$ there are strings which,
given an additional $K$ bits set correctly, obtain a fitness $f +
\delta f$, instead of $f+1$. As a function of $n$, $K$, $q$ and $M$ we
can roughly determine the minimal fitness differential $\delta f$ for
these strings to be preferentially selected. We find that
\begin{equation}
\delta f \geq \frac{n}{(1-q)^K} \left( \frac{1}{\sqrt{M}} + 1 -
  (1-q)^K \right).
\end{equation}
Below this fitness differential, strings of fitness $f+\delta f$ are
effectively neutral with respect to strings with fitness $n$. The net
result is that the parameters of the search, such as $q$ and $M$,
determine a coarse graining of fitness levels where strings in the
band of fitness between $n$ and $n+\delta f$ are treated as having
equal fitness.

In future work we explore how this coarse graining can be turned to
good use by a GA for fitness functions that possess many shallow local
optima---optima that on a coarser scale disappear so that the
resulting coarse-grained fitness function induces a neutral network
architecture like that explored here. Intuitively, it should be
possible to tune GA parameters so that local optima disappear
below the minimal fitness differentials $\delta f$ and so that the GA
efficiently searches the coarse-grained landscape without becoming
pinned to local optima.

\section*{Acknowledgments}

This work was supported at the Santa Fe Institute by NSF Grant
IRI-9705830, ONR grant N00014-95-1-0524, and Sandia National Laboratory
contract AU-4978.

\bibliography{epev}
\bibliographystyle{plain}

\appendix
\section{Selection Operator}
\label{SelectionOperator}

Since the GA uses fitness-proportionate selection, the
proportion $P^s_i$ of strings with fitness $i$ after selection is
proportional to $i$ and to the fraction $P_i$ of strings with
fitness $i$ before selection; that is, $P^s_i = c \; i \; P_i$.
The constant $c$ can be determined by demanding that the distribution
remains normalized. Since the average fitness $\langle f \rangle$ of
the population is given by Eq. (\ref{AvFitness}), we have
$P^s_i = i P_i / {\langle f \rangle}$. In this way, we define a
(diagonal) operator ${\bf S}$ that works on a fitness distribution
$\vec{P}$ and produces the fitness distribution $\vec{P}^s$ after
selection by:
\begin{equation}
\left({\bf S} \cdot \vec{P}\right)_i = \sum_{j=1}^{N+1} \frac{ \delta_{ij}
  j}{\langle f \rangle} P_j ~.
\end{equation}
Notice that this operator is nonlinear since the average fitness
$\langle f\rangle$ is a function of the distribution $\vec{P}$ on
which the operator acts.

\section{Mutation Operator}
\label{MutationOperator}

The component ${\bf M}_{ij}$ of the mutation operator gives the
probability that a string of fitness $j$ is turned into a string with
fitness $i$ under mutation.

First, consider the components ${\bf M}_{ij}$ with $i < j$. These
strings are obtained if mutation leaves the first $i-1$ blocks of
the string unaltered and disrupts the $i$th block in the string.
Multiplying the probabilities that the preceding $i-1$ blocks remain
aligned and that the $i$th block becomes unaligned we have:
\begin{equation}
{\bf M}_{ij} = (1-q)^{(i-1)K} \left( 1 - (1-q)^K\right), \; \; i < j ~.
\end{equation}

The diagonal components ${\bf M}_{jj}$ are obtained when mutation
leaves the first $j-1$ blocks unaltered and does {\em not} mutate
the $j$th block to be aligned. The maximum entropy assumption says
that the $j$th block is random and so the probability $P_a$ that
mutation will change the unaligned $j$th block to an aligned block is
given by:
\begin{equation}
P_a = \frac{1-(1-q)^K}{2^K-1} ~.
\end{equation}
This is the probability that at least one mutation will occur in
the block times the probability that the mutated block will be in the
correct configuration. Thus, the diagonal components are given by:
\begin{equation}
{\bf M}_{jj} = (1-q)^{(j-1)K} (1 - P_a ).
\label{DiagonalMutation}
\end{equation}

Finally, we calculate the probabilities for increasing-fitness
transitions ${\bf M}_{ij}$ with $i > j$. These transition
probabilities depend on the states of the unaligned blocks $j-1$
through $i$. Under the maximum entropy assumption all these blocks are
random. The $j$th block is equally likely to be in any of $2^K-1$
unaligned configurations. All succeeding blocks are equally likely to
be in any one of the $2^K$ configurations, including the aligned one.
In order for a transition to occur from state $j$ to $i$, all the first
$j-1$ aligned blocks have to remain aligned, then the $j$th block
has to become aligned through the mutation. The latter has probability
$P_a$. Furthermore, the following $i-j-1$ blocks all have to be
aligned. Finally, the $i$th block has to be unaligned. Putting these
together, we find that:
\begin{eqnarray}
{\bf M}_{ij} & = & (1-q)^{(j-1)K} P_a
  \nonumber \\
  & & \left( \frac{1}{2^K}\right)^{i-j-1}
  \left( 1 - \frac{1}{2^K} \right) ~, ~ i > j ~.
\end{eqnarray}
The last factor does not appear for the special case of the global
optimum, $i = N+1$, since there is no $(N+1)$st block.

\section{Epoch Fitnesses and Quasispecies}
\label{EpochFitnessQuasispecies}

The generation operator ${\bf G}$ is given by the product of the
mutation and selection operators derived above; i.e. ${\bf G} = {\bf
  M} \cdot {\bf S}$. The operators ${\bf G}^n$ are defined as the
projection of the operator ${\bf G}$ onto the first $n$ dimensions of
the fitness distribution space. Formally:
\begin{equation}
{\bf G}^n_i [ \vec{P} ] ={\bf G}_i [ \vec{P} ] ,\; \; i \leq n,
\end{equation}
and, of course, the components with $i > n$ are zero.

The epoch quasispecies $\vec{P}^n$ is a fixed point of the operator
${\bf G}^n$. As in Sec. \ref{an_appr}, we take out the factor
$\langle f \rangle$ to obtain the matrix ${\bf \tilde{G}}^n$. The
epoch quasispecies is now simply the principal eigenvector of the
matrix ${\bf \tilde{G}}^n$ and this can be easily obtained numerically.

However, in order to obtain analytically the form of the quasispecies
distribution $\vec{P}^n$ during epoch $n$ we approximate the matrix
${\bf \tilde{G}}^n$. As shown in App. \ref{MutationOperator}, the
components ${\bf M}_{ij}$ (and so of ${\bf \tilde{G}}^n$) naturally
fall into three categories. Those with $i < j$, those with $i > j$,
and those on the diagonal $i=j$. Components with $i > j$ involve at
least one block becoming aligned through mutation. These terms are
generally much smaller than the terms that only involve the
destruction of aligned blocks or for which there is no change in the
blocks. We therefore approximate ${\bf \tilde{G}}^n$ by neglecting
terms proportional to the rate of aligned-block creation---what was
called $P_a$ in App. \ref{MutationOperator}. Under this approximation
for the components of ${\bf \tilde{G}}^n$, we have:
\begin{equation}
{\bf \tilde{G}}^n_{ij} = j (1-q)^{(i-1)K} (1 - (1-q)^K) , \; \; i < j ~,
\end{equation}
and 
\begin{equation}
{\bf \tilde{G}}^n_{jj} = j (1-q)^{(j-1) K} ~.
\end{equation}
The components with $i > j$ are now all zero.

Note first that all components of ${\bf \tilde{G}}^n$ only depend on
$q$ and $K$ through $\lambda \equiv (1-q)^K$, the probability that an
aligned block is not destroyed by mutation. Note further that in this
approximation ${\bf \tilde{G}}^n$ is upper triangular. As is well known
in matrix theory, the eigenvalues of an upper triangular matrix are
given by its diagonal components. Therefore, the average fitness
$f_n$ in epoch $n$, which is given by the largest eigenvalue, is equal
to the largest diagonal component ${\bf \tilde{G}}^n$. That is,
\begin{equation}
\label{fn_expression}
f_n = n(1-q)^{(n-1)K} = n \lambda^{n-1} ~.
\end{equation}

The principal eigenvector $\vec{P}^n$ is the solution of the
equation:
\begin{equation}
\sum_{j=1}^n \left( {\bf \tilde{G}}^n_{ij} - f_n \delta_{ij}\right) P^n_j = 0 ~.
\end{equation}
Since the components of ${\bf \tilde{G}}^n$ depend on $\lambda$ in
such a simple way, we can analytically solve for this eigenvector;
finding that the quasispecies components are given by:
\begin{equation}
P^n_i = \frac{(1-\lambda) n \lambda^{n-1-i}}{n \lambda^{n-1-i} - i}
\prod_{j=1}^{i-1} \frac{n \lambda^{n-j} - j}{n \lambda^{n-1-j} -j} ~.
\end{equation} 
For the component $P^n_n$ this reduces to
\begin{equation}
P^n_n =
\prod_{j=1}^{n-1} \frac{n \lambda^{n-j} - j}{n \lambda^{n-1-j} -j} ~.
\label{pnn_expression}
\end{equation}
The above equation can be re-expressed in terms of the epoch fitness
levels $f_j$:
\begin{equation}
P^n_n =
\lambda^{n-1} \prod_{j=1}^{n-1} \frac{f_n- f_j}{f_n - \lambda f_j} ~.
\end{equation}

\end{multicols}

\end{document}